\documentclass[letterpaper,11pt]{article}
  
\usepackage{abstract} 

\usepackage{titlesec} 
\usepackage{authblk} 
\usepackage{datetime} 
\newdateformat{usvardate}{
  \monthname[\THEMONTH] \ordinal{DAY}, \THEYEAR}

  \usepackage[bottom]{footmisc} 

\usepackage[ruled, linesnumbered]{algorithm2e}

\usepackage{fancyhdr}
\fancyheadoffset{0cm}
\usepackage{amssymb}
\usepackage{amsthm}
\usepackage{newtxmath}

\usepackage{microtype} 
\usepackage[utf8]{inputenc}
\usepackage{newtxtext} 
  
\usepackage{setspace} 
\usepackage{lineno,xcolor} 	

\usepackage[top=1.5in, bottom=1.5in, left=1in, right=1in]{geometry}

\usepackage[colorinlistoftodos]{todonotes} 
\usepackage{soul} 
    
\usepackage{graphicx} 
\usepackage{float} 

\usepackage{printlen}

\usepackage{longtable} 
\usepackage{tabularx} 
\usepackage{relsize} 
\usepackage{booktabs}
    
\usepackage{hyperref} 

\begin{document}

\thispagestyle{empty} 


  \title{\vspace{-15mm}\textbf{Pulse of the Pandemic: Iterative Topic Filtering for Clinical Information Extraction from Social Media}}

  \author[1]{\large Julia Wu\footnote{First three authors contributed equally to this research.}}
  \author[2]{\large Venkatesh Sivaraman}
  \author[1]{\large Dheekshita Kumar}
  \author[3]{\large Juan M. Banda}
  \author[1]{\large David Sontag}
  \affil[1]{\normalsize Dept of Electrical Engineering and Computer Science, Massachusetts Institute of Technology}
  \affil[2]{\normalsize Human-Computer Interaction Institute, Carnegie Mellon University}
  \affil[3]{\normalsize Department of Computer Science, Georgia State University}
  \renewcommand\Authands{ and }

  \date{}
	
\maketitle 



\begin{abstract}
    The rapid evolution of the COVID-19 pandemic has underscored the need to quickly disseminate the latest clinical knowledge during a public-health emergency. One surprisingly effective platform for healthcare professionals (HCPs) to share knowledge and experiences from the front lines has been social media (for example, the "\#medtwitter" community on Twitter). However, identifying clinically-relevant content in social media without manual labeling is a challenge because of the sheer volume of irrelevant data. We present an unsupervised, iterative approach to mine clinically relevant information from social media data, which begins by heuristically filtering for HCP-authored texts and incorporates topic modeling and concept extraction with MetaMap. This approach identifies granular topics and tweets with high clinical relevance from a set of about 52 million COVID-19-related tweets from January to mid-June 2020. We also show that because the technique does not require manual labeling, it can be used to identify emerging topics on a week-to-week basis. Our method can aid in future public-health emergencies by facilitating knowledge transfer among healthcare workers in a rapidly-changing information environment, and by providing an efficient and unsupervised way of highlighting potential areas for clinical research.
    
\end{abstract}

\section{Introduction}
In May 2020, a retrospective study of over 3,000 patients in a major New York healthcare system found that in around 1\% of cases, the disease caused by the novel coronavirus (COVID-19) was associated with ischemic stroke \cite{stroke_ny_may}. The result, which was initially described in early April in China and Europe \cite{stroke_wuhan_april,stroke_italy_april} and corroborated by other New York studies \cite{stroke_observational_2, stroke_case_report_2}, quickly became part of a larger story on thrombotic complications of COVID-19 \cite{stroke_dutch_april}. In the weeks leading up to these articles, however, physicians were already discussing a possible association between cerebrovascular accidents and COVID-19 on Twitter, starting with a small number of users affiliated with Boston-area hospitals on March 17. If conversations like these could be surfaced to physicians around the world as they emerge, they could serve as a focal point for new evidence, suggest directions for clinical research, and accelerate progress toward understanding the disease.

In the face of a developing medical situation such as COVID-19, health-care professionals (HCPs) interact with a range of knowledge sources to provide and share up-to-date, accurate clinical information. Guidance released by public health organizations such as the World Health Organization (WHO) and the US Centers for Disease Control and Prevention (CDC) is considered the most reliable, but is relatively slow to change due to its wide impact. These guidelines are backed by published research and case reports, which also take time to be disseminated due to the need for formalization and peer review. To obtain more up-to-the-minute information, HCPs share insights amongst each other through webinars, hospital-specific channels, and chat groups. For example, many initial accounts of Chilblain-like skin lesions on the toes, a now-famous symptom of COVID-19, were circulated on a WhatsApp group from France \cite{covid_toes_history}. 

Social media platforms have emerged as important areas for sharing clinical information publicly. In particular, Twitter is a popular option because it is already home to a sizeable physician community \cite{social_media_physicians_ben_chall,physician_twitter}. HCPs can opt to discuss medical topics under hashtags such as ``\#medtwitter'' and ``\#epitwitter,'' or share experiences tagged as ``free open-access medical education'' (\#FOAMed). While Twitter would not be appropriate as a primary source of clinical guidance, it has been shown to be a useful complement to more incremental, rigorously evaluated sources \cite{twitter_supplement_epi}, and to assist in the dissemination of new information across geographic and cultural boundaries \cite{news_frames_covid_korea}. Furthermore, Twitter data is a well-established source for public health research, and textual analysis of Twitter data can reveal patterns in how health-related content is disseminated through the population \cite{ebola_twitter_network} and identify drug effects and adverse reactions \cite{Sarker2018, Bollegala2018, Cocos2017}. Recently, Wahbeh et al. found themes relating to symptoms and disease transmission in COVID-related tweets by 119 medical professionals \cite{physician_opinion_covid_twitter}, while analysis of a Twitter account compiling anonymous tweets by HCPs highlighted numerous calls to bolster the health system and provide PPE \cite{anonymous_hcps}.

For HCPs looking for clinically-relevant advice or anecdotes, it can be difficult to find the most relevant authors unless they are already in the right social circles. Furthermore, because of the high volume of non-expert conversation, the terms that one would expect to find in clinically-meaningful information can also be found in mundane and non-expert posts as well as in myths and misinformation \cite{covid_misinformation_twitter}. In the case of COVID-19, even within the posts that mention the pandemic, the global impact on everyday life has essentially put ``popularity'' at odds with medical usefulness.

\subsection{Automating clinical relevance filtering}

Framed as an information retrieval problem, our task is to extract and cluster clinically relevant social media posts by reputable authors, who form a tiny minority of the general population. For the purposes of this study, we define information as \textit{clinically relevant} if it is technical in nature and intended to help characterize, prevent, diagnose, or treat the disease under consideration. This definition helps to characterize and remove irrelevant information, especially given the low proportion of clinical tweets compared to noise. While we exclusively consider COVID-19 and HCP-authored Twitter data in this study, the problem formulation and approach can be applied to other healthcare topics or social media platforms.

Prior analyses of social media for public health have taken several approaches that vary in how much supervision is required. One option is to manually label the relevance of a subset of posts, an approach that has shown promise for identifying adverse drug reactions on social media \cite{drug_extraction_classifier,pharmacovigilance_word_embeddings,adverse_drug_reactions_2010}. This is the most straightforward approach, but is difficult to scale and requires expert annotation. Other techniques typically use some form of topic modeling, often based on Latent Dirichlet Allocation (LDA) \cite{LDA}. Since these traditional topic models try to represent \textit{all} posts, including irrelevant ones, some upfront labeling and filtering is still typically required \cite{epidemic-intelligence,avian_flu_relevance_filtering,Kagashe2017}. Hierarchical topic models are another strategy that may facilitate interpretation by connecting related topics \cite{htm-nested,smith2014hiearchie}, enabling individual topics to be more granular. Finally, variants of LDA have been proposed to replace manual document labeling with priors on the topics themselves; for example, SeededLDA can build models around topics for which one or more keywords are known \cite{seededlda}. A recent study by Ferner et al \cite{disaster_seededlda} applies a similar seeded approach with automatically-selected seeds to improve topic quality. However, initial exploration of COVID-19 Twitter data suggested that even these priors may not be strong enough to overcome the bias towards irrelevant topics; we therefore aimed to develop a filtering strategy that could make use of well-established topic modeling tools without any manual labeling.

We take an iterative approach to finding the most clinically relevant documents within a dataset. Documents are automatically annotated for clinical concepts using MetaMap \cite{metamap_citation}, which provides an initial approximation of clinical relevance (though prone to false positives \cite{metamap_error}). Our method uses topic modeling to associate documents with similar content without supervision, then scores topics based on the relevance of the clinical concepts they contain. The documents are then filtered by their degree of association with the most relevant topics, and the process is repeated. The concept relevance estimates are refined in each iteration, thereby overcoming the noise in concept annotations as the filtering quality improves.

We demonstrate the utility of our method by retrospectively analyzing 1 million automatically extracted COVID-related tweets by HCPs, resulting in a detailed picture of clinical discourse about the disease. We perform qualitative and quantitative comparisons of the method against traditional and seeded LDA approaches, and we test the importance of using clinical concepts by comparing to a version of the method that omits the concept extraction step. Finally, we simulate the use of our technique during the early stages of a pandemic by analyzing tweets in two-week time intervals. Our results suggest this method's potential to efficiently surface useful information to a clinical audience without significant manual analysis, potentially before such information is announced in more formal channels.

\section{Methods}

In order to surface clinically relevant information in a highly noisy corpus, we develop a method based on two fundamental subroutines: topic modeling using LDA, and relevance filtering based on clinical concepts. Given an initial corpus of documents (tweets), denoted $\textbf{D}^{(0)}$, we first apply an author-based heuristic (Section \ref{section:heuristic_author_selection}) to obtain a dataset $\textbf{D}^{(1)}$ that has a considerably higher prior likelihood for clinical relevance. Then, we use $\textbf{D}^{(1)}$ to produce a series of filtered subsets $\textbf{D}^{(i)}$ by finding topics, computing a relevance score for each topic, and removing documents containing clinically irrelevant topics. The result of this iterative process (Section \ref{section:iterative_rel_filter}) is a set of highly relevant tweets as well as an interpretable yet granular topic model. This process is summarized in Algorithm \ref{alg1} and detailed in the following sections.

\begin{algorithm}
\SetAlgoLined
 Generate dataset $\mathbf{D}^{(1)}$ from $\mathbf{D}^{(0)}$ \ ;
 \\ Generate MetaMap concepts for all tweets in $\mathbf{D}^{(1)}$ \ ;
 \\ \For{$i=1$ \KwTo $I$}{
    Generate topics using LDA on $\mathbf{D}^{(i)}$ \;
    Score concepts for tweets in $\mathbf{D}^{(i)}$ \;
    Score topics for tweets in $\mathbf{D}^{(i)}$ \;
    Determine threshold for ``relevant topic'' \;
    Use threshold to filter for relevant tweets, generating $\mathbf{D}^{(i+1)}$;
 }
 \caption{Iterative relevance filtering}\label{alg1}
\end{algorithm}

\subsection{Heuristic author selection}
\label{section:heuristic_author_selection}

To generate $\mathbf{D}^{(1)}$, we opt to only consider documents by authors who self-identify as HCPs. Social media norms suggest that it should be relatively straightforward to design a highly sensitive classifier for HCPs: audiences typically rely on author information to determine credibility on Twitter \cite{twitter_bios_credibility}, so users posting about medical topics are incentivized to display their credentials. Thus, we filter $\mathbf{D}^{(0)}$ for users whose name, handle, or bio contains any of 27 medical titles, professions, or keywords (for example, ``MD'', ``Dr'', ``epidemiolog*'', ``public health'').  Note that some authors use credentials falsely or in jest, and many credentialed authors post irrelevant content. This issue could be mitigated by isolating verified accounts, but we intentionally keep non-verified accounts since generally only people with large followings are verified, and this would eliminate several interesting posts from people with smaller followings and those who just began tweeting during the pandemic. We therefore intentionally allow these users' documents to pass the heuristic selection; the subsequent relevance filtering process removes these false positives. Verification of this first heuristic filtering step can be found in section \ref{sec:dataset-info}.

\subsection{Preprocessing and clinical concept extraction}
For each document in the HCP-authored set $\mathbf{D}^{(1)}$, we preprocess the text using standard natural language processing (NLP) routines such as removing contractions, punctuation, HTML tags, and emoji. We lemmatize each word using NLTK's WordNet lemmatizer \cite{nltk}, and remove stopwords and any query terms that were used to generate the original dataset (in our case study data \cite{largedataset}, these were words explicitly referring to the coronavirus).

In addition, we extract clinical concepts from each document with MetaMap18, a tool that uses symbolic NLP to identify UMLS Metathesaurus medical concepts\footnote{In this study, we reduced computational overhead by only applying MetaMap to tweets that contain at least 4 words. Shorter tweets often contained reactions and links, which are not useful for our text analysis purposes.} \cite{metamap_citation}. For a given piece of clinical text, MetaMap outputs a series of “mentions,” or occurrences of a concept defined by a unique identifier (CUI), a preferred name from UMLS, one or more semantic types (e.g. disease/syndrome or clinical finding), and trigger words (a set of words that triggered the concept match). MetaMap also outputs a relevance score, but this was not used in our protocol because it correlated poorly with our desired criteria on Twitter data.

\subsection{Iterative relevance filtering}
\label{section:iterative_rel_filter}
We perform the following iterative procedure to produce corpus $\mathbf{D}^{(i + 1)}$ given $\mathbf{D}^{(i)}$ and $\mathbf{D}^{(i - 1)}$ (for $i \geq 1$).

\subsubsection{Generate topics} 
Using the MALLET implementation of LDA \cite{McCallumMALLET}, we generate a topic model over the $M^{(i)}$ documents in $\textbf{D}^{(i)}$ with $k$ topics. We tested several values of $k$ ranging from 10 to 200, and found that $k = 100$ provides a good balance of detail and summarizability; filtering results are comparable across $k$ values. Topic modeling results in a $k\times M^{(i)}$ matrix $\theta^{(i)}$ that encodes topic probabilities: in particular, the value at $\theta^{(i)}_{t,m}$ is the probability that a word in document $m$ was sampled from topic $t$. As described above, each document is also annotated with a certain number of concepts, which we will denote $C(m)$.

\subsubsection{Score concepts}

Concepts are scored by estimating the clinical relevance of each concept given its trigger word. More formally, we define relevance as a relationship between two corpora $A$ and $B$ where $A \in B$:
    
    \begin{align}
        \text{Rel}(c; A, B) = \frac{f_A(c)/|A| + \epsilon}{(f_B(c)-f_A(c))/(|B| - |A|) + \epsilon}
        \label{eqn:relevance}
    \end{align} \\
where $f_A(c)$ and $f_B(c)$ are the number of occurrences of trigger word $c$ in corpora $A$ and $B$ respectively, $|A|$ and $|B|$ represent the number of documents in each corpus, and $\epsilon$ is a small number to prevent division by zero.

Intuitively, Eqn. \ref{eqn:relevance} measures how frequently concepts appear in preserved documents ($A$) as compared to discarded documents ($B \setminus A$). We assume that the topic-filtration process correctly separates a class of relevant documents (in $A$) from irrelevant documents (in $B$). Thus, a concept's relevance is well-approximated by looking at the ratio of its frequencies in the relevant versus irrelevant corpus. A simpler approach might be to rank concepts by inverse English word frequency, but this tends to artificially elevate alphanumeric codes (often falsely annotated by MetaMap as genes) and suppress everyday words with medical definitions (e.g. ``mask,'' ``vent''). The iterative nature of our protocol affords us the most direct comparison possible, i.e. documents from the same data source that are known to be irrelevant.

The choice of reference set $B$ is flexible: it could be held constant, or it could be set to the previously-generated corpus $\textbf{D}^{(i-1)}$. Empirically, we found that a hybrid of these two approaches results in the most meaningful scores:

\begin{align}
    \text{Rel}^{(i)}(c) =
    \begin{cases}
      \text{Rel}(c; \textbf{D}^{(1)}, \textbf{D}^{(0)}) & \text{if } i = 1\\
      \text{Rel}(c; \textbf{D}^{(i)}, \textbf{D}^{(1)}) & \text{if } i > 1
    \end{cases}    
    \label{eqn:iterative_relevance}
\end{align} \\
In other words, the relevance scores are initialized using the unfiltered set $\textbf{D}^{(0)}$ as reference, comparing HCP to non-HCP texts. In subsequent sections, $\textbf{D}^{(1)}$ serves as a better baseline because of its drastically improved signal-to-noise ratio compared to $\textbf{D}^{(0)}$. We then hold $\textbf{D}^{(1)}$ as a fixed baseline, which helps to stabilize the relevance scores over multiple iterations, and avoids honing in on a particular topic at the expense of others.

We also apply this relevance metric as a pre-filter on the MetaMap concepts: any concepts with $\text{Rel}(c; \textbf{D}^{(1)}, \textbf{D}^{(0)}) < 1$ (the concept occurred less frequently in doctor tweets than non-doctor tweets) are removed. This helps mitigate the signal-to-noise ratio from the very beginning, which clarifies the results of the next step.

\subsubsection{Score topics} Each topic is given a score based on the relevance of the concepts in tweets that are associated with it. Given the document-topic probability matrix $\theta^{(i)}$, the score of topic $t$ is
    
\begin{align}
    \text{Score}^{(i)}(t) = \frac{1}{\sum_{m=1}^{M^{(i)}} \theta^{(i)}_{t,m}} \sum_{m=1}^{M^{(i)}} \theta^{(i)}_{t,m} \sum_{c\in C(m)} \text{Rel}^{(i)}(c) \label{eqn:topic_score}
\end{align} \\
This favors topics for which the documents drawn most heavily from the topic also contain highly relevant concepts. Note that we do not directly test for concept relevance among the topic words $\beta^{(i)}$; this allows for clinically-relevant words that are \textit{not} annotated by MetaMap (e.g. too new to appear in UMLS) to weigh heavily in topics without penalty.

We designate topics as ``relevant'' by choosing a threshold $\tau \in [0, 1]$ and retaining topics that satisfy

\begin{align}
    \text{Score}^{(i)}(t) \geq (S^{(i)}_{\text{max}} - S^{(i)}_{\text{min}}) \cdot \tau + S^{(i)}_{\text{min}}
\end{align}\\ 
where $S^{(i)}_{\text{max}}$ and $S^{(i)}_{\text{min}}$ are the maximal and minimal topic scores, respectively. We denote this set of relevant topics $R^{(i)}$, of size $r^{(i)}$. The threshold $\tau$ reflects the diversity of clinically-relevant topics that is desired; values closer to 1 will tend to keep only the topics with the most clinical terms. We chose $\tau = 0.25$ by examining the distribution of topic scores in an initial 100-topic model, an analysis that can easily be done on a new dataset.

This thresholding scheme allows for variation in how many topics are selected: as the algorithm progresses, the number of topics retained increases with the prevalence of relevant content. We found that after about three rounds of filtering, the number of topics preserved increased significantly, indicating a good stopping point.

\subsubsection{Filter documents}

Finally, we select the documents that are highly associated with relevant topics. We generate the next corpus $\textbf{D}^{(i + 1)}$ by simply choosing tweets in which the probability of sampling from a relevant topic is greater than a uniform probability over topics, i.e. if $\sum_{t \in R^{(i)}} \theta^{(i)}_{t, m} \geq r^{(i)} / k$. At this point, the filtering process can be terminated if $r^{(i)}$ is sufficiently high, or the newly-filtered corpus can be passed on to another iteration of topic modeling and relevance filtering.

\section{Results}

First, we describe our COVID-19 tweet dataset, which forms a case study for the use of our method. Then we present the results of the method on this data (Sec. \ref{sec:filtered-topic-models}), a validation of the use of concept extraction (Sec. \ref{sec:filtering-quality}), and a proof-of-concept for analysis on time-limited datasets (Sec. \ref{sec:st-topic-models}).

\subsection{COVID-19 tweets underscore the need for relevance filtering} \label{sec:dataset-info}

We illustrate the use of our method on a publicly-available COVID-19 Twitter dataset \cite{largedataset}, comprising over 420 million tweets (as of June 21, 2020) that contain coronavirus-related keywords such as \texttt{coronavirus}, \texttt{2019nCoV}, and \texttt{covid19}. The dataset is pre-filtered, with  non-English tweets and retweets removed. Tweets from accounts listed as bots or those that tweet more than 400 times in 1 day were also removed. Using this pre-filtered set, we retrieved 52.9 million tweets posted between January 8 and June 21, 2020 using the \texttt{twarc} command line tool. Notably, many HCP-authored tweets were part of longer ``threads,'' or sequences of tweets by the same author, that were not fully covered by $\mathbf{D}^{(0)}$. We decided to expand the dataset in $\textbf{D}^{(1)}$ by including the complete threads, because the missing tweets often appeared to contain useful clinical information despite not explicitly mentioning COVID keywords. The final HCP-authored tweet set contained 990,756 threads, with 1,078,830 total tweets.

We validated the initial HCP filtering step (Section \ref{section:heuristic_author_selection}) by sampling users in the dataset and annotating whether the users represented HCP individuals or organizations. Because HCPs comprise a very small fraction of the larger dataset, we sampled 500 users from the predicted-HCP set to estimate false positives and 500 users from the predicted-non-HCP set to estimate false negatives. Users were manually labeled by the three first authors as HCPs or non-HCPs based on screen name, username, and bio content, with the majority label taken as ground-truth. Since this heuristic filtering step was designed to capture as many HCPs as possible, we expected its negative predictive value (NPV) to be much stronger than its positive predictive value (PPV). Indeed, we observed a relatively low PPV of 58.2\% and a high NPV of 98.6\%. While the low PPV indicates that around 40\% of the initial HCP set is irrelevant, we note that the filter still dramatically enhances the representation of true positives, which improves our topic modeling results. The subsequent relevance filtering process improves the proportion of true positives even further, removing about 87\% of false-positive accounts and resulting in an overall PPV of 91.5\% for identifying HCPs. 

As expected from the above analysis, the combined presence of non-HCP content and irrelevant content by HCPs resulted in tweets that vary dramatically in relevance, as illustrated in Table \ref{tab:example_tweets}. Some tweets introduce useful information, such as tweet (a) in the table. Many others contain no medical insights (tweet (b)). More subtly, however, tweet (c) contains clinical terms, but does not introduce novel information. Because of the preponderance of tweets like (b) and (c), when we attempted a large topic model ($k = 1000$) without relevance filtering, the clinically-relevant topics were vastly outnumbered by irrelevant ones. This validated the need for filtering out tweets with spurious relevance and preferentially surfacing tweets like (a).

\begin{table}[]
    \centering
    \caption{Paraphrased example tweets from the HCP-authored subset of the dataset.}
    \begin{tabular}{rp{4cm}rp{4cm}rp{4cm}}
    \toprule
        (a) & GI symptoms (anorexia, diarrhea, vomiting, or abdominal pain) may be earlier signs of novel \#coronavirus, presenting before respiratory symptoms \#medtwitter &
        (b) & Feeling low, stressed, or anxious? Free programs available for those experiencing \#MentalHealth challenges associated with \#COVID19 or life’s other issues! &
        (c) & We hate the drug! Take it off the market!!! \includegraphics[height=1em]{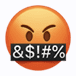} \newline\newline
        *Chloroquine lowers fever in patients with coronavirus*\newline \newline
        Cholorquine: \includegraphics[height=1em]{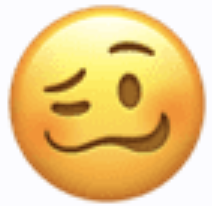}\\
        \bottomrule
    \end{tabular}
    \label{tab:example_tweets}
\end{table}

\subsection{Relevance filtering produces high-quality clinical topics} \label{sec:filtered-topic-models}

We performed three rounds of filtering on the HCP-authored dataset, resulting in a dataset of 107,794 tweets. Roughly 38-40 topics were selected as relevant in each iteration, but 85 topics would have been selected to proceed to the fourth iteration, indicating that most of the irrelevant data had been filtered out by this point.

\begin{figure}
    \centering
    \includegraphics[width=0.9\textwidth]{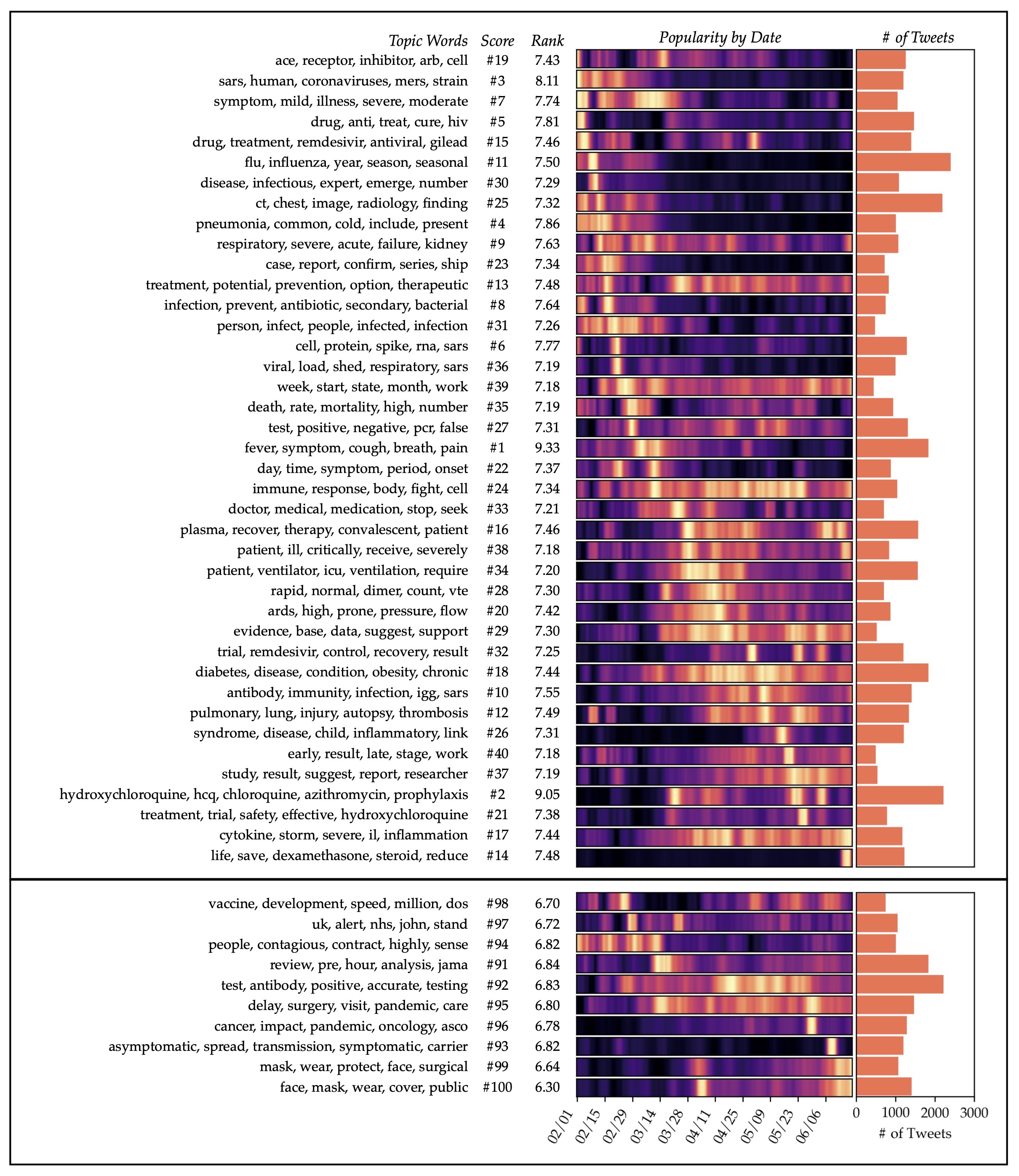}
    \caption{Topics generated after three iterations of relevance filtering (i.e. from $\textbf{D}^{(4)}$). The upper section shows the top 40 highest-scoring topics, while the lower section shows the 10 lowest-scoring topics. Both sections are sorted vertically in order of the date of maximum intensity. The heat map colors indicate the popularity of each topic per day, with yellow representing the peak of popularity for the topic.}
    \label{fig:topic_timeseries}
\end{figure}

The topics extracted from the third level of filtering ($\textbf{D}^{(4)}$), shown in Fig. \ref{fig:topic_timeseries}, demonstrate clear clinical relevance. For example, several topics describe clinical presentations of COVID-19, ranging from the most common symptoms (topic \#1, fever and cough) to rarer manifestations that received buzz among the medical community (\#28, venous thromboembolism; \#26, Kawasaki syndrome). Some topics describe the underlying physiological conditions that lead to these symptoms, such as \#17 (cytokine storms) and \#19 (the ACE2 receptor, which is implicated in viral entry to the cell). Still other topics discuss new and emerging treatments, such as \#16 (convalescent plasma therapy), \#2 (hydroxychloroquine and azithromycin), \#14 (dexamethasone) and \#15 (remdesivir).

The third-iteration model also contains several false positives or irrelevant topics. A few of these are mistakenly scored as relevant, such as \#23 (case reports) and \#29 (supporting evidence); these topics likely co-occur frequently with clinical terms. Topics that are correctly scored as irrelevant often tend to deal with the boilerplate aspects of clinical tweets, such as \#91 (announcing study pre-prints), as well as topics adjacent to COVID-19 but not directly applicable (\#95 and \#96 refer to secondary effects of the pandemic on healthcare). Interestingly, the topics considered least relevant (\#99 and \#100) are the ones that deal with surgical and everyday masks, respectively. These tweets are likely considered irrelevant because they often touch on the U.S. politicization of mask wearing, or address the topic in a public-safety announcement format, such as ``COVID is still spreading out there. Wear your mask. Practice social distancing.'' Irrelevant topics will always be present in some frequency because they co-occur in the same documents as relevant topics, but by presenting the topics in ranked order, the model still facilitates interpretation over an unsorted topic list.

\subsection{Relevance filtering produces higher-quality topics than traditional and seeded LDA}

We validated the performance of our iterative filtering technique by conducting a quantitative and qualitative comparison of topics generated from the final filtered set with those generated using traditional and seeded LDA approaches. These methods were tested on the HCP-authored tweet set, so the prevalence of clinical content was considerably higher than it would be in a random sample of the larger COVID-19 dataset.

For traditional LDA, we used MALLET \cite{McCallumMALLET}, the same implementation used in our pipeline. SeededLDA was tested using the implementation provided by Watanabe and Zhou \cite{seededlda_implementation}. We initialized SeededLDA with two different sets of seeds that would resemble a reasonable expert-derived initialization:

\begin{enumerate}
    \item \textit{Concepts.} Each topic was seeded with one of the most common single-word UMLS concepts in the filtered tweet set. The top 5 concepts were ``lung,'' ``plasma,'' ``pneumonia,'' ``hydroxychloroquine,'' and ``ards.''
    \item \textit{Topics.} Each topic was seeded with the top three words from one of the highest-scoring topics in the relevance-filtered 100-topic model. For example, the top three topics were ``fever, symptom, cough,'' ``hydroxychloroquine, hcq, chloroquine,'' and ``sars, human, coronaviruses.''
\end{enumerate}

\begin{figure}
    \centering
    \includegraphics[width=\textwidth]{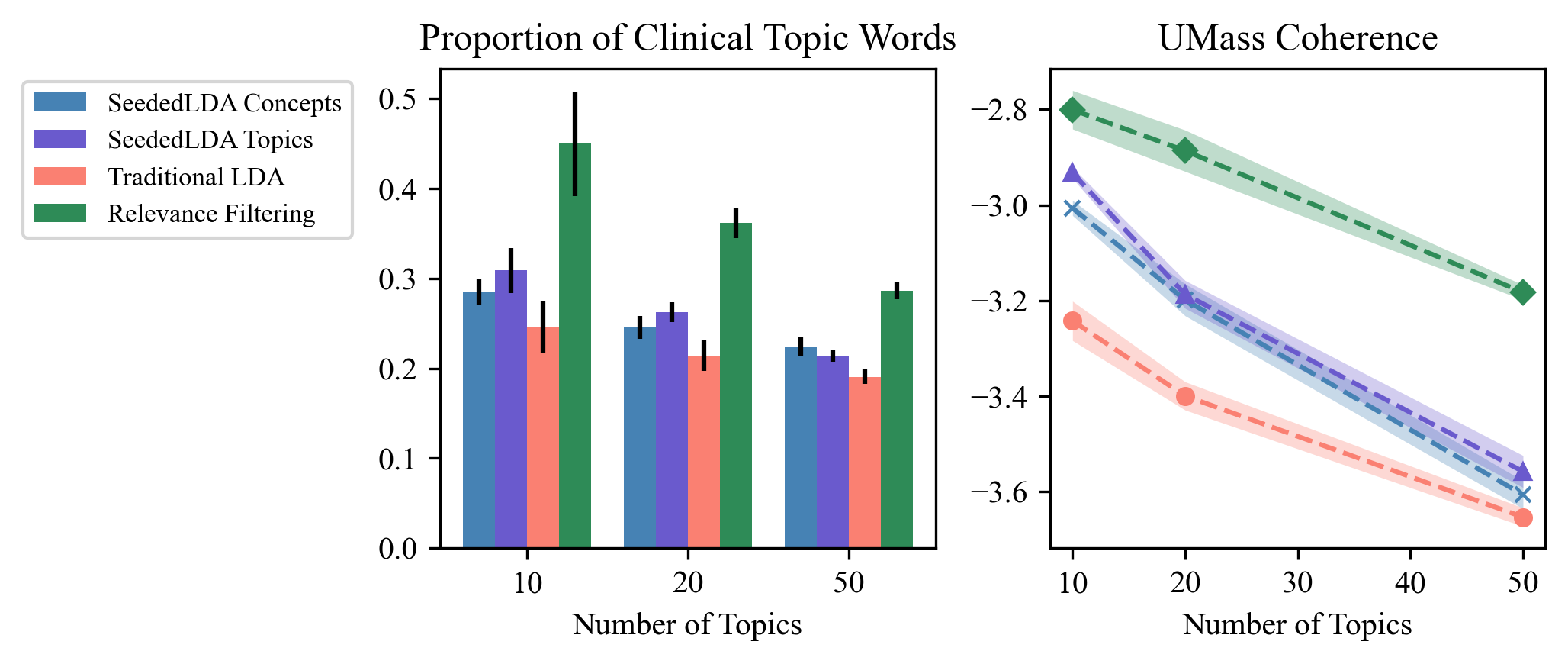}
    \caption{Comparison of the iterative relevance filtering method with traditional and seeded LDA approaches by two metrics, proportion of topic words annotated by MetaMap and the UMass coherence score. Error bars and shaded regions indicate 95\% confidence intervals around the mean over 5 trials.}
    \label{fig:coherence_comparison}
\end{figure}

We quantitatively compared each of these four methods according to two metrics that roughly measure topic quality. The first was the fraction of the top 10 words for each topic that were annotated by MetaMap as a UMLS concept, which approximates the amount of clinical interest in the topic model as a whole. The second was UMass coherence, a well-established metric that aims to correlate with human judgments of topic interpretability \cite{umass_coherence}. Each of the four methods was run 5 times at $k = 10$, $20$, and $50$ topics to compute the two metrics\footnote{For SeededLDA, we initialized the algorithm with $k = 10$, $20$ or $50$ seed topics; the implementation also produces five additional topics for unseeded content, but these were always irrelevant and are excluded from our qualitative analysis.}. The results, shown in Fig. \ref{fig:coherence_comparison}, show that relevance filtering produces more clinically interesting and coherent topics across different values of $k$. Note that relevance filtering should necessarily produce more clinically relevant results than traditional LDA, since our method specifically filters for topics containing clinical concept mentions. However, it also performs better than SeededLDA, even when highly clinically-relevant words are provided to that method as seeds. Similarly, we find that relevance-filtered topics are the most coherent out of the four methods, while SeededLDA's performance is somewhere between traditional LDA and relevance filtering.

The observation of higher topic coherence for relevance filtering was confirmed in a qualitative comparison of the topic words generated by each method. Table \ref{tab:seeded_vs_iterative_topics} shows the top 5 words for each topic in a 20-topic model generated by each method. For SeededLDA, we observe some plausibly meaningful topics, such as ``\textit{lung}, patient, risk, disease, high'' and ``\textit{symptom, illness, mild}, home, stay'' (seed words italicized); however, most of the SeededLDA topics have little to do with their seed words. Several topics share common words in the corpus, such as ``pandemic'' and ``health.'' Notably, the SeededLDA model initialized with top topics from the filtered topic model still failed to associate other clinical terms with the seeds (e.g. the ``cell, spike, protein'' topic was merged with terms about news about UK lockdowns and travel). This suggests a lingering susceptibility to irrelevant content that would need to be mitigated in other ways if SeededLDA were used on this task.

\begin{table}
    \centering \scriptsize \singlespacing
    \begin{tabular}{p{6.2cm}|p{7.2cm}} 
    \toprule
    \normalsize{(a) SeededLDA with Concepts} & \normalsize{(b) SeededLDA with Topics} \\ \midrule
\textit{lung}, patient, risk, disease, high & \textit{symptom}, \textit{cough}, \textit{fever}, day, like \\
\textit{plasma}, late, information, help, new & \textit{hydroxychloroquine}, \textit{hcq}, \textit{chloroquine}, read, article \\
\textit{pneumonia}, care, hospital, patient, doctor & \textit{human}, \textit{sars}, \textit{coronaviruses}, vaccine, china \\
\textit{hydroxychloroquine}, news, read, article, good & \textit{common}, \textit{pneumonia}, \textit{cold}, people, know \\
\textit{ards}, time, life, week, year & \textit{treat}, \textit{anti}, \textit{drug}, trump, american \\
\textit{syndrome}, home, stay, family, day & \textit{cell}, \textit{spike}, \textit{protein}, news, uk \\
\textit{antibodies}, india, lockdown, fight, people & \textit{symptom}, \textit{illness}, \textit{mild}, home, stay \\
\textit{sars}, disease, infection, human, spread & \textit{prevent}, \textit{infection}, \textit{antibiotic}, mask, spread \\
\textit{cough}, mask, social, spread, face & \textit{respiratory}, \textit{severe}, \textit{acute}, care, hospital \\
\textit{fever}, china, country, world, outbreak & \textit{infection}, \textit{antibody}, \textit{immunity}, test, positive \\
\textit{cell}, people, know, dont, like & \textit{year}, \textit{flu}, \textit{influenza}, time, week \\
\textit{dexamethasone}, business, pay, pandemic, job & \textit{lung}, \textit{injury}, \textit{pulmonary}, dr, join \\
\textit{cancer}, dr, join, pandemic, today & \textit{potential}, \textit{treatment}, \textit{prevention}, health, public \\
\textit{treatment}, vaccine, use, study, drug & \textit{life}, \textit{save}, \textit{dexamethasone}, work, thank \\
\textit{azithromycin}, trump, american, president, lie & \textit{treatment}, \textit{drug}, \textit{remdesivir}, patient, study \\
\textit{oxygen}, health, public, government, response & \textit{recover}, \textit{therapy}, \textit{plasma}, india, lockdown \\
\textit{antiviral}, work, thank, help, support & \textit{severe}, \textit{storm}, \textit{cytokine}, business, pay \\
\textit{diabetes}, health, pandemic, people, community & \textit{disease}, \textit{condition}, \textit{diabetes}, people, risk \\
\textit{pcr}, test, positive, contact, people & \textit{ace}, \textit{receptor}, \textit{inhibitor}, state, new \\
\textit{flu}, case, death, new, number & \textit{high}, \textit{ards}, \textit{prone}, case, death
 \\ \midrule
 \normalsize{(c) Traditional LDA} & \normalsize{(d) Iterative Relevance Filtering} \\ \midrule
week, good, dont, thing, year & risk, high, blood, heart, lung \\
open, travel, close, reopen, food & sars, cell, viral, human, ace \\
business, government, job, crisis, pay & people, flu, death, die, rate \\
dr, today, live, question, pm & patient, treatment, evidence, good, life \\
test, positive, contact, symptom, people & patient, pt, icu, ventilator, require \\
vaccine, study, patient, treatment, research & symptom, cough, respiratory, fever, droplet \\
time, child, family, school, feel & patient, health, doctor, hospital, care \\
health, pandemic, public, response, state & immune, response, infection, level, cytokine \\
lockdown, fight, life, india, save & study, data, outcome, mortality, group \\
people, die, life, kill, stop & drug, trial, hydroxychloroquine, treatment, remdesivir \\
case, death, report, number, day & test, antibody, positive, result, negative \\
news, china, world, country, uk & mask, wear, face, protect, public \\
pandemic, community, change, impact, risk & day, case, symptom, infection, report \\
work, support, great, team, pandemic & disease, severe, child, infectious, syndrome \\
read, data, article, important, great & patient, cancer, pandemic, treatment, impact \\
patient, care, hospital, doctor, medical & vaccine, trial, develop, plasma, recover \\
disease, infection, risk, flu, sars & work, great, dr, article, today \\
mask, home, stay, spread, social & patient, pneumonia, clinical, ct, finding \\
late, learn, information, check, free & transmission, asymptomatic, spread, infection, contact \\
trump, american, president, medium, house & people, dont, question, lot, time \\
\bottomrule
    \end{tabular}
    \caption{Comparison of 20-topic models generated by various approaches: (a) SeededLDA seeded with the top 20 most relevant clinical concepts; (b) SeededLDA seeded with the first three words of the 20 highest-scoring topics in Fig. \ref{fig:topic_timeseries}; (c) a traditional MALLET model on the initial HCP-authored tweet corpus; and (d) a MALLET model trained on the final filtered tweet corpus. Words provided as seeds to the SeededLDA algorithm are italicized. Topics are presented in the arbitrary order output by each algorithm.}
    \label{tab:seeded_vs_iterative_topics}
\end{table}

Interestingly, traditional LDA produces several interpretable topics, such as ``case, death, report, number, day'' and ``disease, infection, risk, flu, sars.'' However, these topics are overall less clinically relevant than the model trained on the filtered set, which includes topics like ``patient, pt, icu, ventilator, require'' and ``transmission, asymptomatic, spread, infection, contact.'' Taken together, the quantitative and visual assessments of the four methods indicate that relevance filtering produces the most coherent topics overall, while traditional LDA and SeededLDA are roughly comparable (although SeededLDA tended to merge clinical seed words with unrelated topic words, rendering its results less human-interpretable). Relevance filtering likely eliminated numerous political and economic tweets, leading to the observed improvement in topic quality over traditional LDA.

\subsection{Concept extraction improves focus on clinical terms} \label{sec:filtering-quality}

To validate our use of MetaMap concept annotations when calculating relevance, we ran a version of our iterative filtering routine that entirely omitted clinical concepts. In other words, the relevance of a topic (Eqn. \ref{eqn:topic_score}) was computed not as the sum of relevances over \textit{concepts}, but instead over \textit{all} words in each tweet (excluding stopwords and words explicitly mentioning the coronavirus). This comparison therefore reflects the marginal benefit of directing the relevance filtering toward words already known to be clinically relevant (subject to MetaMap error).

First, we compared the words that were predicted to be most and least relevant by each method, shown in Table \ref{tab:enriched_words}. Relevance is calculated using Eqn. \ref{eqn:iterative_relevance} for all unigrams, bigrams, and trigrams; this therefore also serves as a measure of what phrases are most ``enriched'' in the relevant tweet set compared to the irrelevant tweet set. The first column, which shows the enrichment of $\textbf{D}^{(1)}$ relative to $\textbf{D}^{(0)}$, suggests that heuristic author filtering alone establishes a fairly strong baseline, independent of the use of clinical concepts. Still, after three rounds of relevance filtering, the top phrases in $\textbf{D}^{(4)}$ show a marked improvement over $\textbf{D}^{(1)}$. Using concepts in the filtering process led to an emphasis on \textit{clinical} content (``lung,'' ``respiratory,'' ``ards''), while using all words resulted in a greater proportion of \textit{epidemiological} content (``mortality,'' ``asymptomatic,'' ``growth rate''). This could be because of a natural bias in UMLS toward concepts used in a hospital setting, or because epidemiological topics received the most sustained attention while areas of medical interest shifted. Therefore, while using concepts better suited our objective of extracting clinically-relevant information in this case, omitting the MetaMap step in our pipeline could still result in useful filtering for a different application.

\begin{table}
\centering
\begin{tabular}{r|l|ll} \toprule
 & $\textbf{D}^{(1)}$                & $\textbf{D}^{(4)}$ with concepts & $\textbf{D}^{(4)}$ without concepts \\ \midrule
1                    & medtwitter (\textit{6.33})        & cells (\textit{20.18})             & cells (\textit{28.71})                \\
2                    & publichealth (\textit{4.70})      & lung (\textit{17.67})              & mortality (\textit{28.05})            \\
3                    & physicians (\textit{4.14})        & hcq (\textit{16.03})               & rate (\textit{26.57})                 \\
4                    & patients with (\textit{3.88})     & trial (\textit{15.85})             & asymptomatic (\textit{22.21})         \\
5                    & pts (\textit{3.85})               & patients with (\textit{13.96})     & rate of (\textit{20.66})              \\
6                    & clinical (\textit{3.81})          & severe (\textit{13.64})            & fatality (\textit{18.86})             \\
7                    & physician (\textit{3.66})         & respiratory (\textit{13.63})       & mild (\textit{18.41})                 \\
8                    & icu (\textit{3.44})               & blood (\textit{12.97})             & growth rate (\textit{17.24})          \\
9                    & surgery (\textit{3.34})           & antibodies (\textit{12.46})        & viral (\textit{16.51})                \\
10                   & md (\textit{3.33})                & ards (\textit{12.40})              & ace2 (\textit{16.24})                 \\ \midrule
1                    & f*** (\textit{0.44})              & business (\textit{0.18})           & trump (\textit{0.14})                 \\
2                    & s*** (\textit{0.48})              & leadership (\textit{0.20})         & business (\textit{0.15})              \\
3                    & f***ing (\textit{0.48})           & businesses (\textit{0.23})         & bbc news (\textit{0.19})              \\
4                    & petition (\textit{0.49})          & students (\textit{0.23})           & bbc (\textit{0.20})                   \\
5                    & the petition (\textit{0.50})      & government (\textit{0.23})         & president (\textit{0.21})             \\
6                    & rt (\textit{0.50})                & trump (\textit{0.23})              & news coronavirus (\textit{0.24})      \\
7                    & sign the petition (\textit{0.53}) & crisis (\textit{0.24})             & bbc news coronavirus (\textit{0.25})  \\
8                    & democrats (\textit{0.53})         & county (\textit{0.24})             & the latest (\textit{0.25})            \\
9                    & sign the (\textit{0.53})          & pm (\textit{0.24})                 & businesses (\textit{0.26})            \\
10                   & the petition via (\textit{0.54})  & economy (\textit{0.25})            & amid (\textit{0.26})                 \\ \bottomrule
\end{tabular}
\caption{Words and phrases that were designated most (upper half) and least (lower half) relevant according to Eqn. \ref{eqn:iterative_relevance} (relevance values shown in parentheses). The first column indicates the initial relevance estimates computed after heuristic author filtering; the right two columns list the relevances after three rounds of filtering with and without concept annotations.}
\label{tab:enriched_words}
\end{table}

Next, to ensure our method was accurately discriminating relevant and irrelevant tweets, we examined the effect of filtering on several topics known to be either relevant or irrelevant to the pandemic. With the guidance of a clinician, 12 categories and associated keywords were chosen such that if \textit{any two} of the keywords were present in a tweet, its relevance could be gauged with relative certainty. For example, any tweet containing both ``anosmia'' and ``dysgeusia'' should most likely never be eliminated. Similarly, any tweet containing the names of political leaders is most likely irrelevant. A total of 26,851 tweets were used for this analysis.

\begin{figure}   
\centering
    \includegraphics[width=0.7\textwidth]{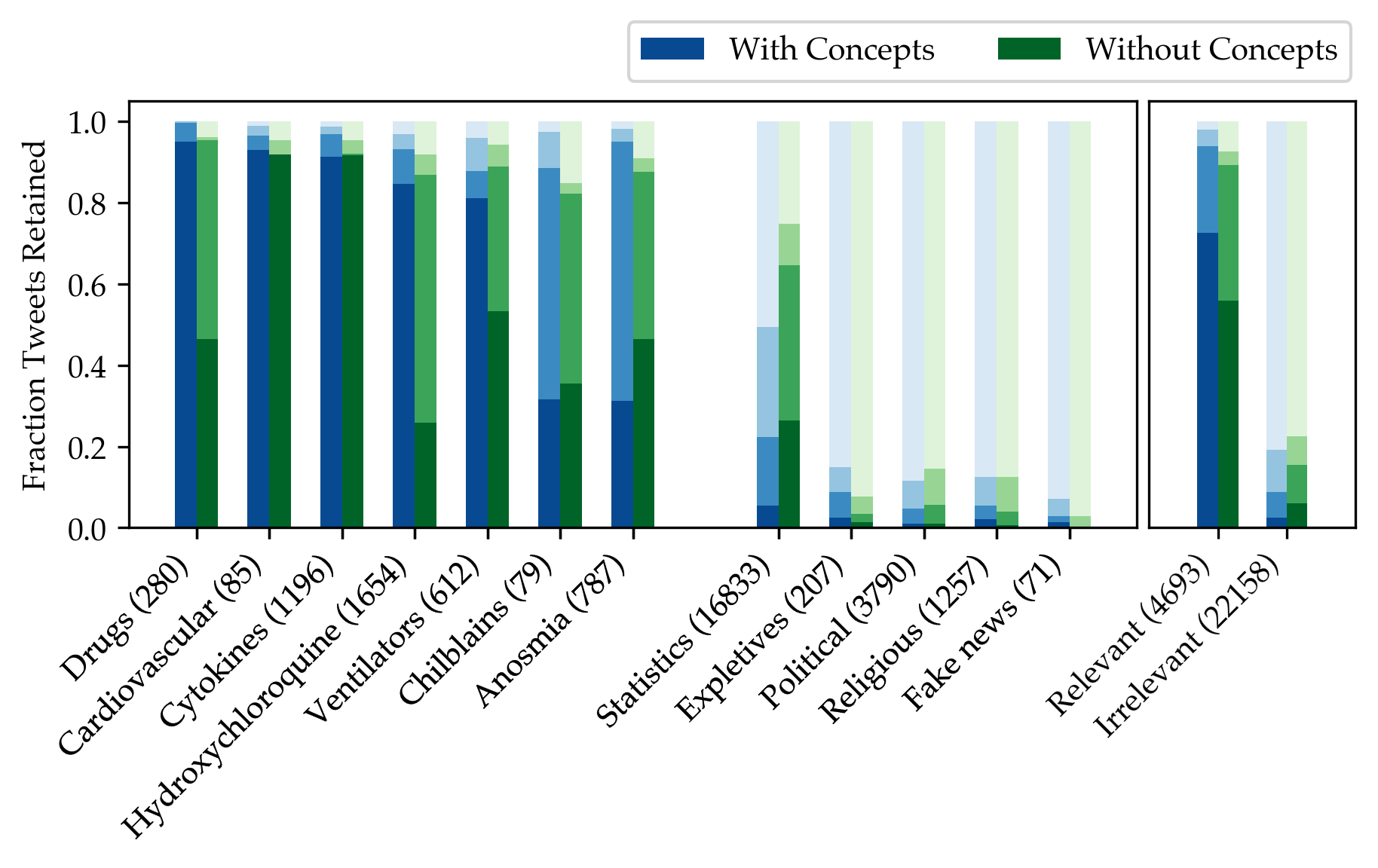}
    \caption{Fraction of tweets preserved by two different models (with and without using clinical concepts for relevance filtering) for each of 12 pre-defined tweet categories, including clinically-relevant subjects (first seven bars) and irrelevant ones (next five). The progressively darker bars represent successive stages of filtering. The parenthesized number indicates the number of tweets that fell into each category. The total fractions for the irrelevant and irrelevant categories are shown in the right panel.}
    \label{fig:tweet_preservation}
\end{figure}

Figure \ref{fig:tweet_preservation} shows the proportion of tweets in each category that were retained by the filtering process in each iteration, with the darkest bar representing the proportion kept after three rounds. Based on the aggregation of categories to the right of the figure, using concept extraction leads to an increase in relevant tweets preserved (72.5\% with concepts, 55.8\% without), and a decrease in irrelevant tweets (2.5\% with concepts, 5.9\% without). Concept filtering performed especially well on the ``Drugs'' category, likely because the drug mentions were consistently annotated and scored well for relevance. On the other hand, concept-based filtering performed poorly by this test on ``Chilblains'' and ``Anosmia,'' both of which are referenced using common words (``toe,'' ``smell'') that were ranked lower by the concept relevance metric than other clinical terms. While the categories used here are imperfect representations of relevance and irrelevance, concept extraction does provide an advantage in preserving the correct sets of tweets.

\subsection{Real-time automatic relevance filtering on Twitter complements academic literature search} \label{sec:st-topic-models}

While the topic models analyzed above were built retrospectively on a dataset containing nearly six months worth of tweets, applying our method during the early stages of a pandemic would require a more real-time approach. We therefore investigated whether applying automatic relevance filtering to time-windowed subsets of the data could recover topics of clinical interest, or micro-trends, during those periods. This experiment also provided a natural opportunity to probe the time-scale differences between Twitter and academic articles, two information sources that evolved at different rates throughout the pandemic. We hypothesized that clinical keywords would be surfaced in topic models soon after their first mention on Twitter, and that they would appear in HCP-authored tweets and clinically-relevant topics \textit{before} they appeared in academic articles.

\begin{figure}
    \centering
    \includegraphics[width=0.75\textwidth]{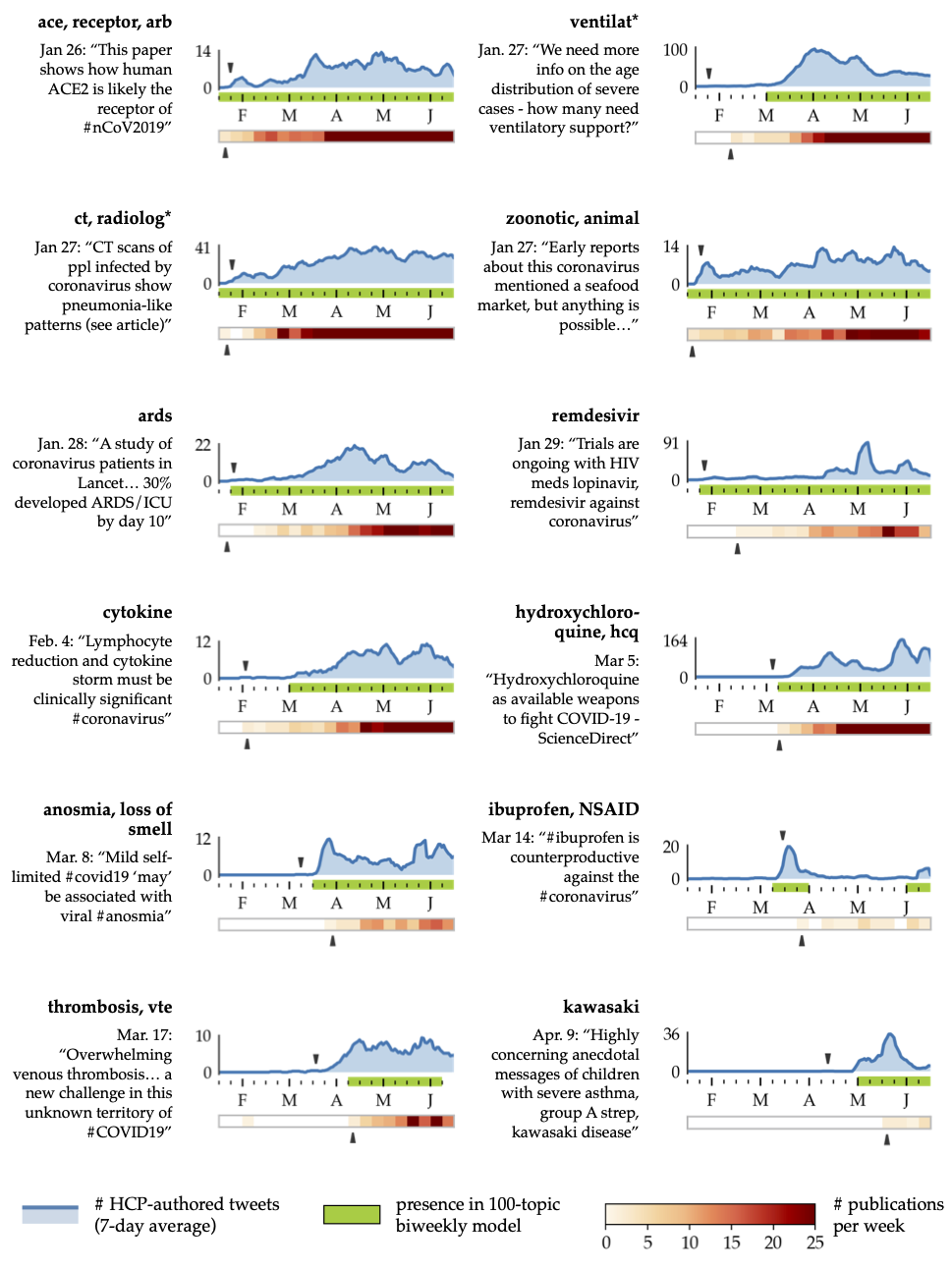}
    \caption{Incidence of selected topics of clinical interest in both Twitter and academic publications. The number of tweets containing the bolded keywords is plotted in blue as a 7-day moving average (note the different $y$-axis scales), while the red heatmap bar shows the number of new publications per week. Time intervals are highlighted in green if the topic keywords appear in the topic model for that interval (specifically, in the top ten words for each topic). The black carets mark the date of the first tweet (paraphrased at left of each plot) and the first publication relevant to the topic.}
    \label{fig:timeline}
\end{figure}

To generate time-windowed topic models, we first split the dataset into twenty 2-week subsets, where the subsets overlapped each other by 1 week. Topic models with $k = 100$ topics were generated after two rounds of relevance filtering on each subset. The resulting topics were highly granular and highlighted new clinical concepts from week to week, similar to the topics shown in Fig. \ref{fig:topic_timeseries}. From these topics, we curated a set of twelve diverse topics that were of known clinical interest and that could be identified fairly unambiguously using a small set of keywords. These keywords were converted to regular expressions (to allow for variations like ``radiology'' and ``radiological''), then used to search both our dataset of HCP-authored tweets and CORD-19, a dataset of COVID-related publications \cite{CORD-dataset}. Fig. \ref{fig:timeline} shows the results of this analysis for each of the twelve topics.

\subsubsection{Comparison with initial tweet appearance}
Many of the keywords appear in the time-windowed topic models in the same week as or shortly after the first tweet that contains them (indicated by the upper black caret). This is a remarkable finding, given that several of these keywords are mentioned in fewer than 20 tweets per week (out of about 9,000 total tweets per week on average). The cases in which topic model appearance is delayed from the first tweet mention, such as \textbf{thrombosis}, generally have even smaller tweet counts; however, as shown below, these tweets could still be found within other clinically-relevant topics. These results support the notion that relevance-filtered topic models can be a fairly sensitive indicator of clinical interest.

\subsubsection{Comparison with academic publications}
Next, we consider the question of whether the time-windowed topic models surface topics of interest in advance of academic publications. Note that this is only possible when the first tweets about a topic appear before the first publication, which occurred for six out of the twelve example topics. Even so, Fig. \ref{fig:timeline} shows that for the most part, the time-windowed topic models surface topics of interest at the same time or prior to their appearance in academic publications. For instance, topics about clinical trials and approvals for the antiviral drug \textbf{remdesivir} appeared in almost all time-windowed topic models from Jan. 26 onward, while the first mention of the drug in CORD-19 was only in mid-February. Another example is the micro-trend discussing the use of \textbf{ibuprofen} to treat COVID-19 symptoms, which became popular and controversial in mid-March and was briefly addressed in a March 25 article. These cases support the usefulness of Twitter topics in surfacing clinical research with long lead times before publication, as well as potentially controversial and rapidly developing micro-trends.

In contrast, we found that publications for the concepts related to \textbf{cytokine} and \textbf{ventilators} predate their first appearance in the time-windowed topic models. This seems to be because the bulk of discussion of these topics in our English-language dataset does not begin until March, when COVID-19 first started to hit English-speaking countries. While early tweets mentioning these topics were present in the dataset, they were typically isolated and served to highlight publications describing the disease in early epicenters like China (e.g. \cite{huang2020clinical} is an article from late January referencing \textbf{ARDS}). This reflects a fundamental drawback to Twitter trend analysis: in order to appear in a topic model, the topic must be actively discussed by HCPs on Twitter, who are in some ways systematically non-representative of the global medical community. Another complication of this analysis is that some concepts (for example, \textbf{ace} and \textbf{zoonotic}) appear in both time-windowed topic models and publications at the very start of January, implying that our dataset only starts amid existing discussion of these topics. 

\begin{figure*}[t!]
    \centering
    \includegraphics[width=0.7\textwidth]{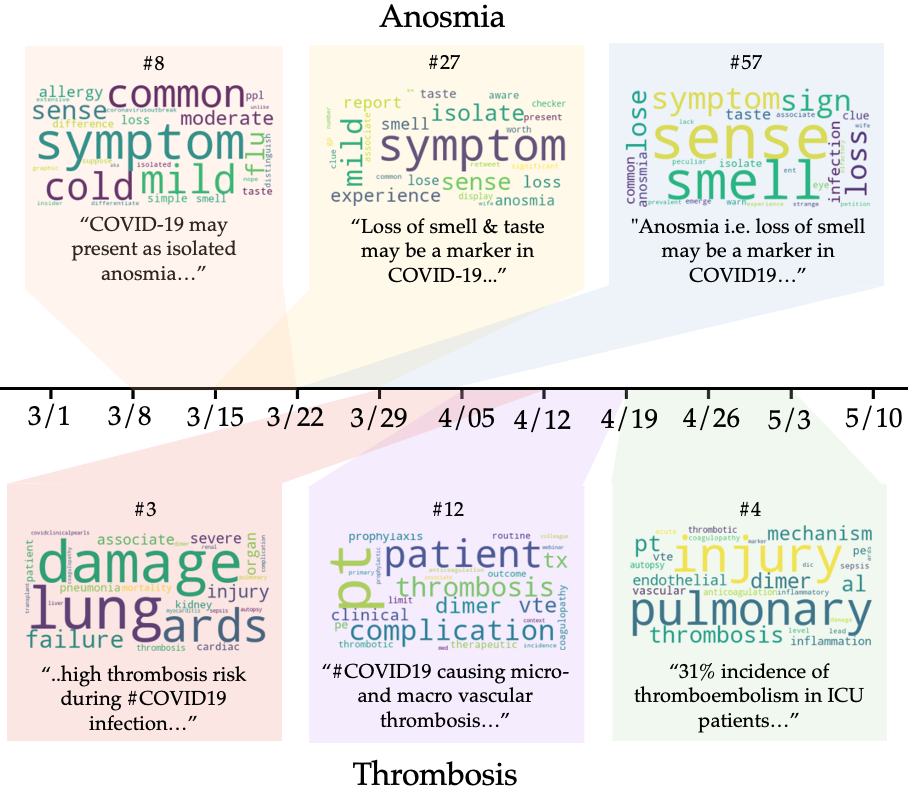}
    \caption{Topics containing the same surfaced clinical concepts plotted across time. Topics containing the concept \textbf{anosmia} from the time intervals 3/8-3/22, 3/15-3/29, and 3/22-4/05 are shown on top. Topics containing the concept \textbf{thrombosis} from the time intervals 3/29-4/12, 4/05-4/19, and 4/19-5/3 are shown at the bottom. Each topic is rendered as a word cloud; the size of the words correlates to its weight in the topic. A paraphrased tweet belonging to each topic along with the topic's clinical relevance ranking relative to other topics in that time interval are also shown.}
    \label{fig:wordclouds}
\end{figure*}

\subsubsection{Evolution of time-windowed topics}
Most of the example topics were represented across several two-week windows, reflecting continued clinical interest after their initial discovery. To illustrate how the content of the topics changed throughout these time intervals, Fig. \ref{fig:wordclouds} illustrates word clouds for topics containing the keywords for two of the example topics, \textbf{anosmia} and \textbf{thrombosis}.

While \textbf{anosmia} is now known to be a symptom of COVID-19, in March 2020 the connection between anosmia and COVID-19 was not yet well established. In topics that contain anosmia-related keywords from three consecutive time windows (Fig. \ref{fig:wordclouds}, top), words describing loss of smell grow progressively more significant, which tracks with the increasing number of tweets about anosmia during this time period. A similar trend can be seen for topics related to \textbf{thrombosis} (blood clotting, Fig. \ref{fig:wordclouds}, bottom), although the keywords are initially mixed together with words related to lung failure. Even so, the tweets highly associated with this early topic contain relevant mentions of thrombosis, underscoring the importance of looking not only at the topic words, but at the texts they are drawn from.

\section{Discussion}

This study provides a new look at social media for rapidly-evolving public health situations, and develops a strategy for extracting granular clinical topics without manual labeling. As with all social media applications, a key challenge in information extraction from Twitter is the signal-to-noise ratio: our filtering process reduced the initial COVID dataset to about 0.2\% of its original size. Furthermore, because of the pandemic's impact on everyday life, many tweets can contain medical terminology without necessarily imparting clinically relevant information. Our proposed technique resolves these issues by applying topic modeling and concept extraction in tandem, resulting in progressively better estimates of clinical relevance.

Traditional topic modeling is a valuable first step in understanding the contents of any textual dataset, and we observed that our topic models often naturally grouped together relevant and irrelevant information. However, the larger and more granular our models became, the harder it was to find relevant topics without a ranking strategy (which led to the development of our method). On the other hand, MetaMap concept extraction can identify relevant phrases regardless of their frequency, a useful complement to statistical NLP methods. Using these annotations to rank the topic models allows one to quickly discard irrelevant categories, improving interpretability considerably for larger models. Concept extraction therefore approximates a physician's prior belief of which topics should be considered \textit{relevant}, while the topic-document mapping approximates which documents should be considered \textit{related}.

An iterative approach is key to enabling topic modeling and concept extraction to refine and validate each other's results. For example, in early iterations, we observed many tweets that were associated with relevant topics because they contained clinical terminology, but that were not themselves written in a medical ``frame.'' As these tweets were carried forward into subsequent levels, however, topics became more relevant overall, and tweets that used clinical terminology in a non-medical context became sequestered into their own topics. We note that the filtered topic sets do contain false positives, but their prevalence dramatically decreases over the course of filtering.

The volume and diversity of clinically-relevant content found in this study is a testament to the opportunity for medical professionals to learn from each other and initiate new lines of research based on topics extracted from Twitter. Our filtered dataset seems to capture micro-trends of special interest to physicians such as emerging symptoms and clinical anecdotes, which augment and enrich an understanding gained from academic articles. In fact, the first tweets in these topics of interest often themselves contained links to publications, including on domain-specific platforms such as AuntMinnie.com (an online community of medical imaging professionals). This was especially the case for early clinical topics, perhaps because they originated in non-English language countries, as well as topics stemming from biological research (e.g. the role of the ACE2 receptor). In later months, the dataset includes more anecdotal discussions that document the English-speaking medical community's evolving understanding of COVID-19. By surfacing both timely references to published findings and records of real-time clinical developments, topic models of HCP-authored tweets could greatly accelerate awareness of new developments outside of the circles in which they are shared.

The relevance filtering strategy presented here suggests future research questions that can take advantage of the scale and high signal-to-noise ratio of data it generates. In particular, many downstream linguistic analyses, such as clinical concept co-occurrence or sentiment analysis, would benefit from a dataset that is highly focused on medical content. Time-limited topic models such as the ones described in Section \ref{sec:st-topic-models} could also be aligned and compared with one another to automatically highlight new clinical indicators. Answering these research questions could help understand the events of 2020 as well as improve preparedness for future public-health emergencies.

As with any study involving Twitter data, our analysis is subject to the biases and caveats inherent in social media. The dataset collected here only consisted of English tweets, but it could be more helpful to apply the technique across languages and capture a truly global conversation, perhaps by integrating a multilingual topic model \cite{multilingual-topic-models}. Even among English-speaking HCPs, this dataset is hardly a representative sample of all medical opinion on the pandemic, since the demographics of physicians on Twitter are inevitably skewed by the demographics of Twitter itself. Finally, the rapid oscillations of interest in the generated topics suggest that HCPs, like other Twitter users, gravitate toward the most exciting stories at any given time. The degree of intensity of a topic, therefore, will always be an imperfect proxy for true clinical importance.

\section{Conclusions}
We present a method that combines topic modeling and clinical concept extraction to surface clinically relevant information from a noisy text dataset. Both topic modeling and concept annotation have their limitations on this type of data; nevertheless, we have shown that iterating between the two techniques can overcome these weaknesses. This is especially important in emergent situations such as COVID-19, where unconventional data sources and unsupervised information extraction are often the only option for rapid analysis. Our pipeline can be applied without modification to social media data on other medical topics, and the iterative filtering technique could be a helpful augmentation to topic models used for preliminary explorations of text data.

In seeking to extract clinical information from social media data, this study addresses a problem space that has not yet been addressed to our knowledge. Indeed, in normal circumstances the rapid fluctuations of social media trends are antithetical to robust, reliable official guidance. As demonstrated by our analysis, however, HCP-authored tweets about COVID-19 can present a surprisingly bountiful window into the epicenters of a pandemic, and we hope that our method enables future research to gain deeper insights from the physician population on social media. By automatically bringing clinical tweets out of the limited audiences among whom they are shared, newly-emerging clinical observations can be disseminated quickly and clinicians everywhere can contribute to shared knowledge. Just as Twitter data reflects the rapidly-changing world, this method could enable the flexible, real-time analysis that a pandemic demands.

\section*{Acknowledgements}

The authors would like to thank Dr. Jonathan Bickel (Boston Children's Hospital, Harvard Medical School) for providing invaluable support and guidance throughout the project. The authors would also like to thank Matthew McDermott for his mentorship and encouragement. Finally, the authors are grateful to the anonymous reviewers for providing valuable feedback toward improving the manuscript. This work received funding from a grant by the National Institute of Aging (3P30AG059307-02S1).

\singlespacing


    \bibliographystyle{abbrv}
    {\footnotesize
	\bibliography{biblio}} 

\begin{thebibliography}{10}

\bibitem{metamap_citation}
A.~R. Aronson and F.-M. Lang.
\newblock An overview of metamap: historical perspective and recent advances.
\newblock {\em Journal of the American Medical Informatics Association},
  17(3):229–236, 2010.

\bibitem{largedataset}
J.~M. Banda, R.~Tekumalla, G.~Wang, J.~Yu, T.~Liu, Y.~Ding, and G.~Chowell.
\newblock {A Twitter Dataset of 100+ million tweets related to COVID-19}, Mar.
  2020.
\newblock {This dataset will be updated bi-weekly at least with additional
  tweets, look at the github repo for these updates}.

\bibitem{nltk}
S.~Bird, E.~Loper, and E.~Klein.
\newblock {\em Natural Language Processing with Python}.
\newblock O'Reilly Media Inc.

\bibitem{htm-nested}
D.~M. Blei, M.~I. Jordan, T.~L. Griffiths, and J.~B. Tenenbaum.
\newblock Hierarchical topic models and the nested chinese restaurant process.
\newblock In {\em Proceedings of the 16th International Conference on Neural
  Information Processing Systems}, NIPS’03, page 17–24, Cambridge, MA, USA,
  2003. MIT Press.

\bibitem{LDA}
D.~M. Blei, A.~Y. Ng, and M.~I. Jordan.
\newblock Latent dirichlet allocation.
\newblock {\em J. Mach. Learn. Res.}, 3(null):993–1022, Mar. 2003.

\bibitem{Bollegala2018}
D.~Bollegala, S.~Maskell, R.~Sloane, J.~Hajne, and M.~Pirmohamed.
\newblock Causality patterns for detecting adverse drug reactions from social
  media: Text mining approach.
\newblock {\em {JMIR} Public Health and Surveillance}, 4(2):e51, May 2018.

\bibitem{multilingual-topic-models}
J.~Boyd-Graber and D.~Blei.
\newblock Multilingual topic models for unaligned text, 2012.

\bibitem{Cocos2017}
A.~Cocos, A.~G. Fiks, and A.~J. Masino.
\newblock Deep learning for pharmacovigilance: recurrent neural network
  architectures for labeling adverse drug reactions in twitter posts.
\newblock {\em Journal of the American Medical Informatics Association},
  24(4):813--821, Feb. 2017.

\bibitem{epidemic-intelligence}
E.~Diaz-Aviles, A.~Stewart, E.~Velasco, K.~Denecke, and W.~Nejdl.
\newblock Epidemic intelligence for the crowd, by the crowd (full version),
  2012.

\bibitem{disaster_seededlda}
C.~Ferner, C.~Havas, E.~Birnbacher, S.~Wegenkittl, and B.~Resch.
\newblock Automated seeded latent dirichlet allocation for social media based
  event detection and mapping.
\newblock {\em Information}, 11(8), 2020.

\bibitem{metamap_error}
D.~A. Hanauer, M.~Saeed, K.~Zheng, Q.~Mei, K.~Shedden, A.~R. Aronson, and
  N.~Ramakrishnan.
\newblock Applying metamap to medline for identifying novel associations in a
  large clinical dataset: a feasibility analysis.
\newblock {\em Journal of the American Medical Informatics Association :
  JAMIA}, 21(5):925—937, 2014.

\bibitem{huang2020clinical}
C.~Huang, Y.~Wang, X.~Li, L.~Ren, J.~Zhao, Y.~Hu, L.~Zhang, G.~Fan, J.~Xu,
  X.~Gu, et~al.
\newblock Clinical features of patients infected with 2019 novel coronavirus in
  {Wuhan}, {China}.
\newblock {\em The Lancet}, 395(10223):497--506, 2020.

\bibitem{seededlda}
J.~Jagarlamudi, H.~Daum{\'e}~III, and R.~Udupa.
\newblock Incorporating lexical priors into topic models.
\newblock In {\em Proceedings of the 13th Conference of the {E}uropean Chapter
  of the Association for Computational Linguistics}, pages 204--213, Avignon,
  France, Apr. 2012. Association for Computational Linguistics.

\bibitem{Kagashe2017}
I.~Kagashe, Z.~Yan, and I.~Suheryani.
\newblock Enhancing seasonal influenza surveillance: Topic analysis of widely
  used medicinal drugs using twitter data.
\newblock {\em Journal of Medical Internet Research}, 19(9):e315, Sept. 2017.

\bibitem{stroke_dutch_april}
F.~Klok, M.~Kruip, N.~van~der Meer, M.~Arbous, D.~Gommers, K.~Kant, F.~Kaptein,
  J.~van Paassen, M.~Stals, and M.~e.~a. Huisman.
\newblock Incidence of thrombotic complications in critically ill icu patients
  with covid-19.
\newblock {\em Thrombosis Research}, 191:145--147, 2020.

\bibitem{adverse_drug_reactions_2010}
R.~Leaman, L.~Wojtulewicz, R.~Sullivan, A.~Skariah, J.~Yang, and G.~Gonzalez.
\newblock Towards internet-age pharmacovigilance: extracting adverse drug
  reactions from user posts in health-related social networks.
\newblock In {\em Proceedings of the 2010 workshop on biomedical natural
  language processing}, pages 117--125, 2010.

\bibitem{stroke_observational_2}
Y.~Li, M.~Li, M.~Wang, Y.~Zhou, J.~Chang, Y.~Xian, D.~Wang, L.~Mao, H.~Jin, and
  B.~Hu.
\newblock Acute cerebrovascular disease following covid-19: a single center,
  retrospective, observational study.
\newblock {\em Stroke and Vascular Neurology}, 2020.

\bibitem{ebola_twitter_network}
H.~Liang, I.~C.-H. Fung, Z.~T.~H. Tse, J.~Yin, C.-H. Chan, L.~E. Pechta, B.~J.
  Smith, R.~D. Marquez-Lameda, M.~I. Meltzer, K.~M. Lubell, et~al.
\newblock How did ebola information spread on twitter: broadcasting or viral
  spreading?
\newblock {\em BMC public health}, 19(1):438, 2019.

\bibitem{drug_extraction_classifier}
X.~Liu and H.~Chen.
\newblock Azdrugminer: An information extraction system for mining
  patient-reported adverse drug events in online patient forums.
\newblock In {\em Smart Health - International Conference, ICSH 2013,
  Proceedings}, Lecture Notes in Computer Science (including subseries Lecture
  Notes in Artificial Intelligence and Lecture Notes in Bioinformatics), pages
  134--150, Aug. 2013.
\newblock 2013 International Conference for Smart Health, ICSH 2013 ;
  Conference date: 03-08-2013 Through 04-08-2013.

\bibitem{stroke_italy_april}
C.~Lodigiani, G.~Iapichino, L.~Carenzo, M.~Cecconi, P.~Ferrazzi, T.~Sebastian,
  N.~Kucher, J.-D. Studt, C.~Sacco, B.~Alexia, and et~al.
\newblock Venous and arterial thromboembolic complications in covid-19 patients
  admitted to an academic hospital in milan, italy.
\newblock {\em Thrombosis Research}, 191:9–14, 2020.

\bibitem{stroke_wuhan_april}
L.~Mao, H.~Jin, M.~Wang, Y.~Hu, S.~Chen, Q.~He, J.~Chang, C.~Hong, Y.~Zhou,
  D.~Wang, X.~Miao, Y.~Li, and B.~Hu.
\newblock {Neurologic Manifestations of Hospitalized Patients With Coronavirus
  Disease 2019 in Wuhan, China}.
\newblock {\em JAMA Neurology}, 77(6):683--690, 06 2020.

\bibitem{covid_toes_history}
P.~R. Massey and K.~M. Jones.
\newblock Going viral: A brief history of chilblain-like skin lesions (“covid
  toes”) amidst the covid-19 pandemic.
\newblock {\em Seminars in Oncology}, 2020.

\bibitem{McCallumMALLET}
A.~K. McCallum.
\newblock Mallet: A machine learning for language toolkit.
\newblock http://mallet.cs.umass.edu, 2002.

\bibitem{umass_coherence}
D.~Mimno, H.~M. Wallach, E.~Talley, M.~Leenders, and A.~McCallum.
\newblock Optimizing semantic coherence in topic models.
\newblock In {\em Proceedings of the Conference on Empirical Methods in Natural
  Language Processing}, EMNLP '11, page 262–272, USA, 2011. Association for
  Computational Linguistics.

\bibitem{physician_twitter}
R.~Mishori, L.~O. Singh, B.~Levy, and C.~Newport.
\newblock Mapping physician twitter networks: Describing how they work as a
  first step in understanding connectivity, information flow, and message
  diffusion.
\newblock {\em J Med Internet Res}, 16(4):e107, Apr 2014.

\bibitem{twitter_bios_credibility}
M.~R. Morris, S.~Counts, A.~Roseway, A.~Hoff, and J.~Schwarz.
\newblock Tweeting is believing? understanding microblog credibility
  perceptions.
\newblock In {\em Proceedings of the ACM 2012 Conference on Computer Supported
  Cooperative Work}, CSCW ’12, page 441–450, New York, NY, USA, 2012.
  Association for Computing Machinery.

\bibitem{pharmacovigilance_word_embeddings}
A.~Nikfarjam, A.~Sarker, K.~O’Connor, R.~Ginn, and G.~Gonzalez.
\newblock {Pharmacovigilance from social media: mining adverse drug reaction
  mentions using sequence labeling with word embedding cluster features}.
\newblock {\em Journal of the American Medical Informatics Association},
  22(3):671--681, 03 2015.

\bibitem{stroke_case_report_2}
T.~J. Oxley, J.~Mocco, S.~Majidi, C.~P. Kellner, H.~Shoirah, I.~P. Singh, R.~A.
  De~Leacy, T.~Shigematsu, T.~R. Ladner, K.~A. Yaeger, M.~Skliut,
  J.~Weinberger, N.~S. Dangayach, J.~B. Bederson, S.~Tuhrim, and J.~T. Fifi.
\newblock Large-vessel stroke as a presenting feature of covid-19 in the young.
\newblock {\em New England Journal of Medicine}, 382(20):e60, 2020.
\newblock PMID: 32343504.

\bibitem{social_media_physicians_ben_chall}
S.~Panahi, J.~Watson, and H.~Partridge.
\newblock Social media and physicians: Exploring the benefits and challenges.
\newblock {\em Health Informatics Journal}, 22(2):99--112, 2014.

\bibitem{news_frames_covid_korea}
H.~W. Park, S.~Park, and M.~Chong.
\newblock Conversations and medical news frames on twitter: Infodemiological
  study on covid-19 in south korea.
\newblock {\em J Med Internet Res}, 22(5):e18897, May 2020.

\bibitem{Sarker2018}
A.~Sarker, M.~Belousov, J.~Friedrichs, K.~Hakala, S.~Kiritchenko, F.~Mehryary,
  S.~Han, T.~Tran, A.~Rios, R.~Kavuluru, B.~de~Bruijn, F.~Ginter, D.~Mahata,
  S.~M. Mohammad, G.~Nenadic, and G.~Gonzalez-Hernandez.
\newblock Data and systems for medication-related text classification and
  concept normalization from twitter: insights from the social media mining for
  health ({SMM}4h)-2017 shared task.
\newblock {\em Journal of the American Medical Informatics Association},
  25(10):1274--1283, Oct. 2018.

\bibitem{covid_misinformation_twitter}
L.~Singh, S.~Bansal, L.~Bode, C.~Budak, G.~Chi, K.~Kawintiranon, C.~Padden,
  R.~Vanarsdall, E.~Vraga, and Y.~Wang.
\newblock A first look at covid-19 information and misinformation sharing on
  twitter, 2020.

\bibitem{smith2014hiearchie}
A.~Smith, T.~Hawes, and M.~Myers.
\newblock Hiearchie: Visualization for hierarchical topic models.
\newblock In {\em Proceedings of the Workshop on Interactive Language Learning,
  Visualization, and Interfaces}, pages 71--78, 2014.

\bibitem{anonymous_hcps}
K.~J. Sullivan, M.~B. MD, A.~K. MSPH, J.~M. Banda, and L.~E. Hunter.
\newblock Characterization of anonymous physician perspectives on covid-19
  using social media data.
\newblock {\em Biocomputing 2021}, pages 95--106.

\bibitem{physician_opinion_covid_twitter}
A.~Wahbeh, T.~Nasralah, M.~Al-Ramahi, and O.~El-Gayar.
\newblock Mining physicians' opinions on social media to obtain insights into
  covid-19: Mixed methods analysis.
\newblock {\em JMIR Public Health Surveill}, 6(2):e19276, Jun 2020.

\bibitem{CORD-dataset}
L.~L. Wang, K.~Lo, Y.~Chandrasekhar, R.~Reas, J.~Yang, D.~Eide, K.~Funk, R.~M.
  Kinney, Z.~Liu, W.~Merrill, P.~Mooney, D.~Murdick, D.~Rishi, J.~Sheehan,
  Z.~Shen, B.~Stilson, A.~Wade, K.~Wang, C.~Wilhelm, B.~Xie, D.~Raymond, D.~S.
  Weld, O.~Etzioni, and S.~Kohlmeier.
\newblock Cord-19: The covid-19 open research dataset.
\newblock {\em ArXiv}, 2020.

\bibitem{seededlda_implementation}
K.~Watanabe and Y.~Zhou.
\newblock Theory-driven analysis of large corpora: Semisupervised topic
  classification of the un speeches.
\newblock {\em Social Science Computer Review}, 0(0):0894439320907027, 0.

\bibitem{stroke_ny_may}
S.~Yaghi, K.~Ishida, J.~Torres, B.~M. Grory, E.~Raz, K.~Humbert, N.~Henninger,
  T.~Trivedi, K.~Lillemoe, S.~Alam, M.~Sanger, S.~Kim, E.~Scher,
  S.~Dehkharghani, M.~Wachs, O.~Tanweer, F.~Volpicelli, B.~Bosworth, A.~Lord,
  and J.~Frontera.
\newblock Sars-cov-2 and stroke in a new york healthcare system.
\newblock {\em Stroke}, 51(7):2002--2011, 2020.

\bibitem{twitter_supplement_epi}
Y.~T. Yang, M.~Horneffer, and N.~DiLisio.
\newblock Mining social media and web searches for disease detection.
\newblock {\em Journal of Public Health Research}, 2(1):e4, May 2013.

\bibitem{avian_flu_relevance_filtering}
S.~Yousefinaghani, R.~Dara, Z.~Poljak, T.~M. Bernardo, and S.~Sharif.
\newblock The assessment of twitter’s potential for outbreak detection: Avian
  influenza case study.
\newblock {\em Scientific Reports}, 9(1), 2019.

\end{thebibliography}


\end{document}